\documentclass[apl,superscriptaddress,twocolumn,showpacs,amsmath,amssymb]{revtex4-1}

\usepackage{graphicx}
\usepackage{bm}
\usepackage[breaklinks=true,colorlinks,citecolor=blue,linkcolor=blue,urlcolor=blue]{hyperref}
\usepackage{color}
\usepackage{amsmath}
\usepackage{datetime}
\usepackage{gensymb}
\usepackage[normalem]{ulem}

\begin{document}
\title{Thermoelectric conversion at $30\,{\rm K}$ in InAs/InP nanowire quantum dots} 
\author{Domenic Prete}
\affiliation{NEST, Scuola Normale Superiore and Istituto Nanoscienze-CNR, Piazza S. Silvestro 12, I-56127 Pisa, Italy}
\author{Paolo Andrea Erdman}
\affiliation{NEST, Scuola Normale Superiore and Istituto Nanoscienze-CNR, Piazza S. Silvestro 12, I-56127 Pisa, Italy}
\author{Valeria Demontis}
\affiliation{NEST, Scuola Normale Superiore and Istituto Nanoscienze-CNR, Piazza S. Silvestro 12, I-56127 Pisa, Italy}
\author{Valentina Zannier}
\affiliation{NEST, Scuola Normale Superiore and Istituto Nanoscienze-CNR, Piazza S. Silvestro 12, I-56127 Pisa, Italy}
\author{Daniele Ercolani}
\affiliation{NEST, Scuola Normale Superiore and Istituto Nanoscienze-CNR, Piazza S. Silvestro 12, I-56127 Pisa, Italy}
\author{Lucia Sorba}
\affiliation{NEST, Scuola Normale Superiore and Istituto Nanoscienze-CNR, Piazza S. Silvestro 12, I-56127 Pisa, Italy}
\author{Fabio Beltram}
\author{Francesco Rossella}
\affiliation{NEST, Scuola Normale Superiore and Istituto Nanoscienze-CNR, Piazza S. Silvestro 12, I-56127 Pisa, Italy}
\author{Fabio Taddei}
\affiliation{NEST, Scuola Normale Superiore and Istituto Nanoscienze-CNR, Piazza S. Silvestro 12, I-56127 Pisa, Italy}
\author{Stefano Roddaro}
\affiliation{NEST, Scuola Normale Superiore and Istituto Nanoscienze-CNR, Piazza S. Silvestro 12, I-56127 Pisa, Italy}
\affiliation{Dipartimento di Fisica ``E. Fermi'', Universit\`a di Pisa, Largo Pontecorvo 3, I-56127 Pisa, Italy}

\begin{abstract}
We demonstrate high-temperature thermoelectric conversion in InAs/InP nanowire quantum dots by taking advantage of their strong electronic confinement. The electrical conductance $G$ and the thermopower $S$ are obtained from charge transport measurements and accurately reproduced with a theoretical model accounting for the multi-level structure of the quantum dot. Notably, our analysis does not rely on the estimate of co-tunnelling contributions since electronic thermal transport is dominated by multi-level heat transport. By taking into account two spin-degenerate energy levels we are able to evaluate the electronic thermal conductance $K$ and investigate the evolution of the electronic figure of merit $ZT$ as a function of the quantum dot configuration and demonstrate $ZT\approx35$ at $30\,{\rm K}$, corresponding to an electronic efficiency at maximum power close to the Curzon-Ahlborn limit.
\end{abstract}
	
\maketitle

The progress in the fabrication and control of nanostructured systems has opened new prospects for thermoelectric (TE) research~\cite{Hicks93,Mahan96} and has provided new ways to create improved TE devices~\cite{Dresselhaus07,Vineis10,Shi12,Sothmann2015,Benenti2017}.
In particular, quantum dots (QDs) were soon identified as ideal systems for the implementation of efficient heat engines~\cite{Mahan96} and for the creation of nanoscale thermometers~\cite{Hoffmann2007}. The key property is their discrete density of states that yields a strong energy selectivity in their transmission profile thus opening the way to the realization of TE systems with an optimized performance~\cite{Mahan96}. As a consequence, TE effects in QDs have been extensively investigated both theoretically~\cite{BeenakkerTE92,Dzurak1998,Boese2001,Andreev2001,Turek2002,Koch2002,Kubala2006,Zianni2007,Kubala2008,Zianni2008,Jacquet2009,Costi2010,Billings2010,Mani11,Entin2012,Rejec2012,Jordan2013,Lopez2013,Kennes2013,Dutt2013,Muralidharan2013,Sierra2014,Svilans2016,Zimbovskaya2016} and experimentally~\cite{Dzurak1993,Staring93,Molenkamp1994,Dzurak1997,Scheibner2007,Svensson11,Svensson13,Josefsson2018,Feshchenko2014,Dutta17,Dutta2018}.
The TE behavior of electronic devices is characterized by the so-called figure of merit $ZT$ that is larger in more efficient devices. Indeed, within linear-response regime, both the maximum efficiency and the efficiency at maximum power are solely determied by the figure of merit, and they are growing functions of ZT~\cite{Benenti2017}. The figure of merit is defined as $ZT=GS^2T/K$, where $G$ is the electrical conductance, $S$ is the Seebeck coefficient, $K$ is the thermal conductance and $T$ is the temperature. In bulk materials the maximization of $ZT$ has been proved to be highly nontrivial since $G$, $S$ and $K$ are intrinsically related. On the contrary, heat and charge transport in QD-based TE devices can be disentangled and larger values of $ZT$ can be achieved. The performance of thermal machines based on QDs has been theoretically studied by many authors \cite{Mani11,Erdman17,Zianni2010,Sanchez2011,Karlstroem2011,Trocha2012,Jordan2013}, while only one experimental investigation is available up to now~\cite{Josefsson2018}. We also note that QDs can provide a key building block for the fundamental investigation of quantum and stochastic thermodynamic effects, as shown for small-sized systems, where thermal fluctuations are of much relevance~\cite{Blickle2011,pekola2015,Ronage2016}.
 
The investigation of TE effects in single-electron systems is performed in temperature regimes for which the thermal energy $k_{\rm B}T$ is smaller than the Coulomb gap. In addition, it is often desirable to implement devices where the individual energy levels are well-resolved with respect to the thermal energy so that only few levels contribute to the heat transport.
Here, we exploit the strong confinement of QDs realized in InAs/InP heterostructured nanowires (NWs) \cite{Bjoerk2004,Romeo2012,Rossella2014} to demonstrate field-effect control on the thermopower $S$ up to temperatures of the order of $30\,{\rm K}$. The high temperature regime has been little explored in similar systems \cite{Svensson13}, usually operated at much lower temperatures~\cite{Svensson11,Josefsson2018}. Our devices allow for the application of a temperature gradient along the NW, the measurement of charge current and the Seebeck coefficient, as well as the direct measurement of local temperatures. The strong confinement of the electrons in our QDs allows us to obtain charging energies $E_{\rm c}$ exceeding $10\,{\rm meV}$, with inter-level spacings $\Delta \varepsilon$ of the order of $5\,{\rm meV}$ and thus to operate our devices at relatively high temperatures and close to the QD depletion. Using a theoretical model accounting for sequential tunnelling mediated by two spin-degenerate levels \cite{Beenakker1991Cond,Zianni2008,Erdman17} we can accurately reproduce the experimental conductance and thermopower. This allows us to estimate the electronic thermal conductance $K$, the power factor $GS^2$ and the electronic figure of merit $ZT$.

\begin{figure}[tb!]
	\centering
		\includegraphics[width=\linewidth]{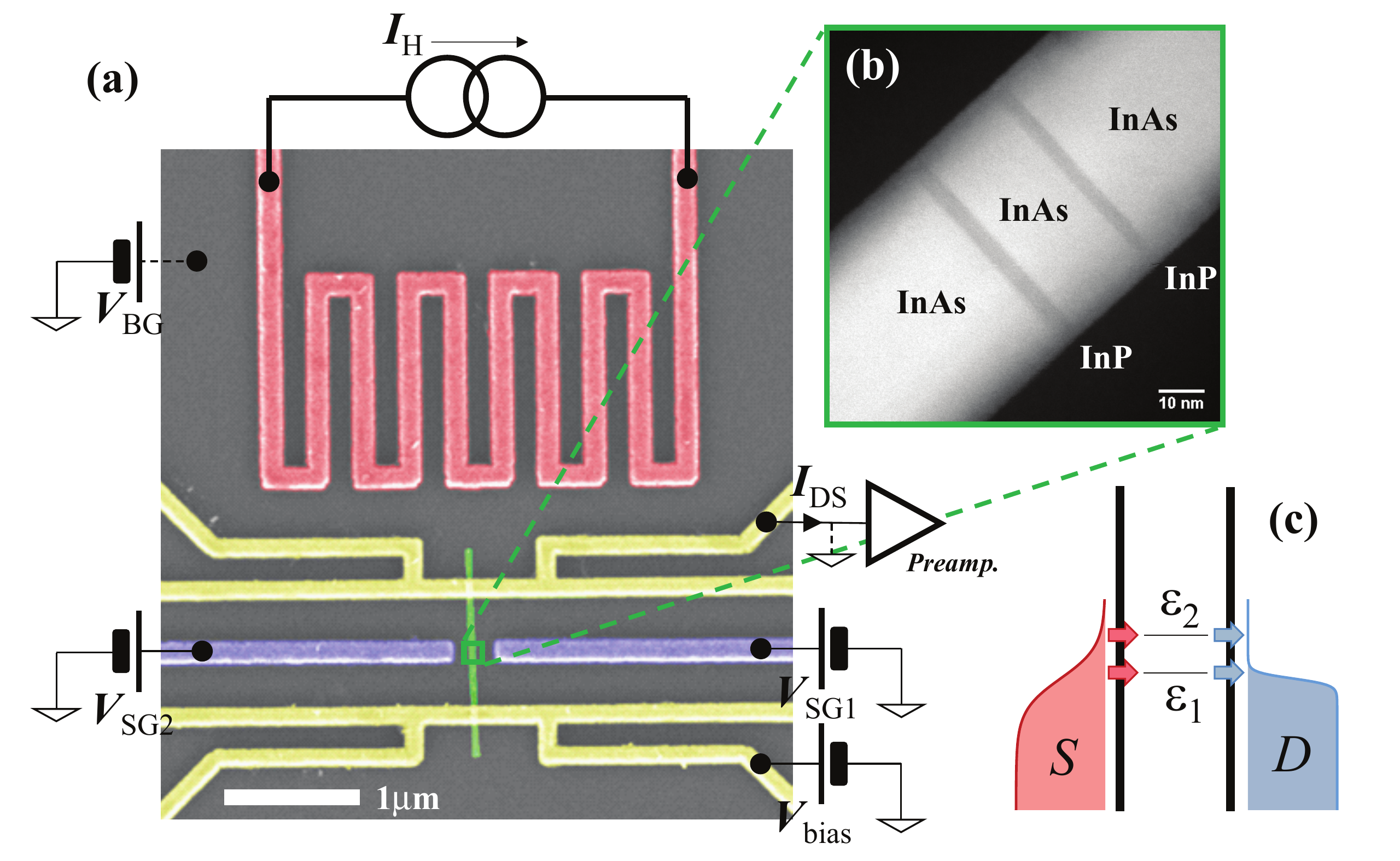}
	\caption {(a) Scanning electron micrograph of a typical device. A local heater (red) is used to establish a temperature difference $\Delta T$ between the two ends of the NW (green) embedding an InAs/InP heterostructured QD. The device is fabricated on top of a degenerately-doped SiO$_2$/Si p++ substrate (gray) and a set of Ti/Au electrodes (yellow) can be used both as electrical contacts to the NW and as local resistive thermometers. The QD electronic configuration can be controlled with a pair of side gates ($SG$, purple) or using the conductive substrate as a backgate electrode ($BG$). (b) Transmission electron micrograph of the heterostructured QDs. (c) Sketch of the energetics scheme: the QD implements a multi-level system that can mediate heat and charge transport between the a source (S) and a drain (D) electrodes, in the presence of thermal and electric biases. Two spin-degenerate levels $\varepsilon_1$ and $\varepsilon_2$ ($\Delta\varepsilon=\varepsilon_2-\varepsilon_1$) play a relevant role in the regime studied in the experiment.}
	\label{fig1}
\end{figure}

Fig. \ref{fig1}(a) shows a scanning electron micrograph of one of the investigated devices, where different colors correspond to different elements. The device core is constituted by an InAs/InP heterostructured NW with a diameter of $52\pm 1\,\rm nm$. As visible in the Fig. \ref{fig1}(b), the nanostructure embeds a $20\pm\,1\,\rm nm$ InAs island separated from the rest of the NW by two $\approx 4\, \rm nm$-wide InP barriers. The NWs used for this study were grown by Au-seeded chemical beam epitaxy~\cite{Zannier2018} and have a wurtzite crystal structure. As grown NWs were detached from the growth substrate by sonication in isopropyl alcohol (IPA) and randomly deposited on a SiO$_2$/Si p++ substrate by drop-casting and contacted by e-beam lithography followed by an evaporation of a metallic Ti/Au ($10/100\, \rm nm$) bilayer. The resulting device layout includes a metallic serpentine heater (red), which can be fed by a current $I_{\rm H}$ to induce, thanks to Joule heating, a thermal gradient along the NW (green). Our architecture also includes a set of multiple contact electrodes (yellow) that allows to: (i) apply a voltage bias $V_{\rm bias}$ and measure the current $I_{\rm DS}$ flowing through the NW; (ii) perform a local measurement of the temperature by tracking the resistance of the central metallic part of the contact electrodes. The QD population and spectrum can be controlled by a set of field-effect gates including the SiO$_2$/Si p++ substrate, or back-gate (gray), and two side gates that can also be used to modify the radial confinement profile of the QD (purple) \cite{rod11}.

The control on the temperature of the system is obtained by setting the temperature of the bath $T_{\rm bath}$ in which the device is immersed and then by setting a temperature gradient across the NW so that its hot end is at a temperature $T_{\rm H}$ and its cold end is at a temperature $T_{\rm C}$.
The average temperature $T_{\rm avg}=(T_{\rm H}+T_{\rm C})/2$ is set high enough ($T_{\rm avg}\sim30\,\rm K$) so that we can benefit of the following advantages.
(i) Since $\hbar\Gamma \ll k_BT_{\rm avg}$ (where $\Gamma$ is the characteristic tunnelling rate through the QD barriers) and $\Delta\varepsilon$ is not much larger than $k_{\rm B}T_{\rm avg}$, sequential tunnelling processes dominate the thermoelectric transport and data analysis is relatively straightforward (see below for details);
(ii) Since $\Delta T\ll T_{\rm avg}$, non-linear response effects in thermal bias $\Delta T$ are negligible thus simplifying the theoretical analysis; (iii) charge transport can be explored for relatively large applied bias voltages (up to $V_{\rm bias}\approx k_{\rm B}T_{\rm avg}/e$) while remaining in the linear response regime. In this situation, the current $I_{\rm DS}$ flowing through the QD is given by

\begin{figure}[tb!]
	\centering
	\includegraphics[width=\linewidth]{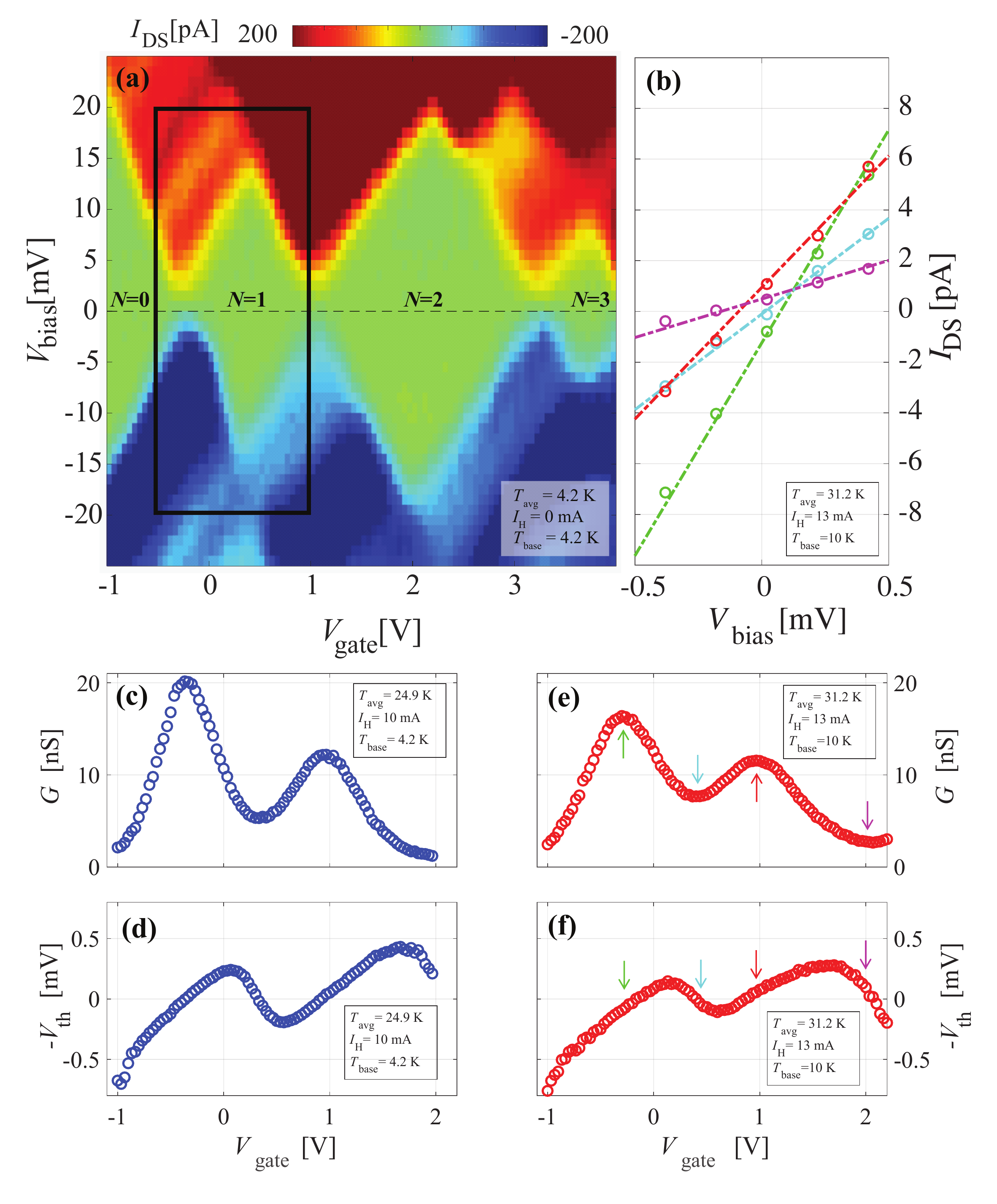}
	\caption{(a) Typical map of the QD current versus $V_{\rm bias}$ and $V_{\rm gate}$ showing the QD filling N in each Coulomb blockade diamond. The analysis of the diamonds yields a charging energy $E_{\rm c}=14\,\pm\,2\,\rm{meV}$ and an energy spacing between the first and second QD level $\Delta\varepsilon=5\,\pm\,1\,\rm{meV}$. (b) $I_{\rm DS}(V_{\rm bias})$ curves for several values of the gate voltage $V_{\rm gate}$. Thermoelectric measurements are obtained in the linear response regime, i.e. for low applied voltage bias ($\rm eV_{\rm bias}<\rm K_BT$) for which the response of the system is linear and thus eq. \ref{eq_lin} holds.(b) Current $I_{\rm DS}$, as a function of $V_{\rm bias}$, in the linear response range for four values of $V_{\rm gate}$ highlighted by arrows of the corresponding color in panels (e) and (f). (c) Extrapolated conductance $G$ and (d) thermovoltage $-V_{\rm th}$ as function of the applied gate voltage $V_{\rm gate}$ for the first degenerate energy level for an average temperature $T_{\rm avg}=24.9\,{\rm K}$ and for $I_{\rm H}=10\,\rm mA$ current feeding the heating serpentine, when the system is in a $T_{\rm bath}=\rm 4.2\,\rm K$ thermal bath. Panels (e) and (f) report equivalent data for $T_{\rm avg}=31.2\,\rm K$ and $I_{\rm H}=13\,\rm mA$ and $T_{\rm bath}=10\,\rm K$. Temperatures are obtained by using the local metallic thermometers at the two NW ends.}
	\label{fig2}
\end{figure}

\begin{equation}
\label{eq_lin}
I_{\rm DS}=G(V_{\rm bias}+S\Delta T),
\end{equation} 

\noindent where $G$ is the electrical conductance and $S$ is the Seebeck coefficient. The thermovoltage at open circuit conditions is thus given by $V_{\rm th}=-S\Delta T$. It should be noted that both $G$ and $S$ are temperature dependent even if not explicitly indicated. Typical transport data from our QDs can be seen in Fig.~\ref{fig2}(a), reporting a colorplot of $I_{\rm DS}$ as a function of the gate voltage $V_{\rm gate}$ and the DS bias $V_{\rm bias}$ at $T=4.2\,{\rm K}$. Any of the three available gates (SG1, SG2 and BG) can be used to operate the device, but in the experiment reported here the QD was controlled using the lateral gate SG1 because it yielded the most stable control of the electronic configuration. Coulomb diamonds are clearly visible for various QD filling numbers $N$, and indicate a charging energy on the first spin-degenerate level $E_{\rm c}=14\,\pm\,2\,{\rm meV}$ and an inter-level spacing $\Delta\varepsilon=5\,\pm\,1\,{\rm meV}$. In Fig. \ref{fig2}(a) $N=0$ corresponds to a completely depleted QD, as further discussed in the Supplementary Information and as indicated by high-temperature transport data. However, the behavior of the device is not expected to critically depend on the presence of filled levels as long as they are far away in energy with respect to the relevant energy scales, in particular $k_{\rm B}T_{\rm avg}$. In Fig. \ref{fig2}(b) we report a set of $I_{\rm DS}(V_{\rm bias})$ curves corresponding to vertical cross-sections at fixed $V_{\rm gate}$ taken from the region highlighted by the black rectangle in panel (a). The thermoelectric response of the device has been obtained from transport data in the linear regime. From the curve $I_{\rm DS}(V_{\rm bias})$ curve restricted to the linear response range (see e.g. Fig 2 (b)), we determine $G$ (slope) and $I_{\rm th}=GS\Delta T$ (intercept) for each $V_{\rm gate}$ value and for a fixed $\Delta T$ using Eq.~(\ref{eq_lin}). From the knowledge of the slope and intercept of the lines it is possible to compute $V_{\rm th}=-S\Delta T$. Fig.~\ref{fig2}(c)-(f) show the plots of $G(V_{\rm gate})$ and $-V_{\rm th}(V_{\rm gate})$ for two different bath temperatures and heating currents $I_{\rm H}$, such that the average temperatures of the QD are $T_{\rm avg}=24.9\,\rm K$ in panels (c) and (e) and $T_{\rm avg}=31.2\,\rm K$ in panels (d) and (f), respectively.

Experimental data are analyzed with a theoretical model based on a master equation accounting for sequential tunneling of electrons between the leads and the QD~\cite{BeenakkerTE92,Nazarov2009,Erdman17}. As already mentioned in the introduction, we can reasonably assume that co-tunneling processes are negligible when two conditions are met, namely $\hbar\Gamma \ll k_BT_{\rm avg}$ and $\Delta\varepsilon \lesssim k_BT_{\rm avg}$. The first condition is required since small values of $\hbar\Gamma/ k_BT_{\rm avg}$ suppress the coefficient of co-tunneling rates. The second condition comes from the fact that co-tunneling contributions decay as a power law with the energy difference $\delta E$ between the chemical potential of the leads and the nearest resonance (measured in units of $k_BT_{\rm avg}$)~\cite{Nazarov2009}. This means that they may prevail over sequential contributions for large enough $\delta E$, since sequential tunneling contributions decay exponentially with $\delta E/k_BT_{\rm avg}$~\cite{Beenakker1991Cond,Averin1992,Nazarov2009}. Remarkably, the presence of two spin-degenerate levels, with spacing equal to $\Delta\varepsilon$, sets a bound to the energy difference $\delta E$. The second condition (for which $\Delta\varepsilon \lesssim k_BT_{\rm avg}$), therefore, ensures that the exponential suppression of sequential tunneling is limited and we can safely disregard co-tunneling effects. This is particularly relevant for values of $V_{\rm gate}$ such that the chemical potential of the leads is between the two spin-degenerate levels, where the thermopower is maximal even though the conductance is negligible.

Two spin-degenerate QD orbitals $\varepsilon_1$ and $\varepsilon_2$ are taken into account in the QD model and tunneling rates are assumed to depend on the number of electrons $N$ in the QD (see Supplementary Information for details about the model). Further important outputs of the fitting procedure are the electronic temperatures $T_{\rm H,fit}$ and  $T_{\rm C,fit}$ that can be used to obtain a better estimate of the net temperature bias across the QD structure $\Delta T_{\rm fit}$. The average fit temperature $T_{\rm avg,fit}=(T_{\rm H,fit}+T_{\rm C,fit})/2$ matches nicely with the one obtained experimentally sing the metallic thermometers $T_{\rm avg}$, and confirms the good calibration of the resistive thermometers. On the contrary, $\Delta T_{\rm fit}$ is sizably smaller than $\Delta T$. This is not very surprising since the NW has a finite heat conductance leading to a partitioning of the thermal bias as observed in experiments at lower temperatures~\cite{Hoffmann09}. Since $\Delta T_{\rm fit}$ is the relevant temperature bias for the analysis of the TE response of the QD, it has been used for all the subsequent estimates on the TE parameters of the nanostructure.

\begin{figure}[tb!]
	\centering
	\includegraphics[width=\linewidth]{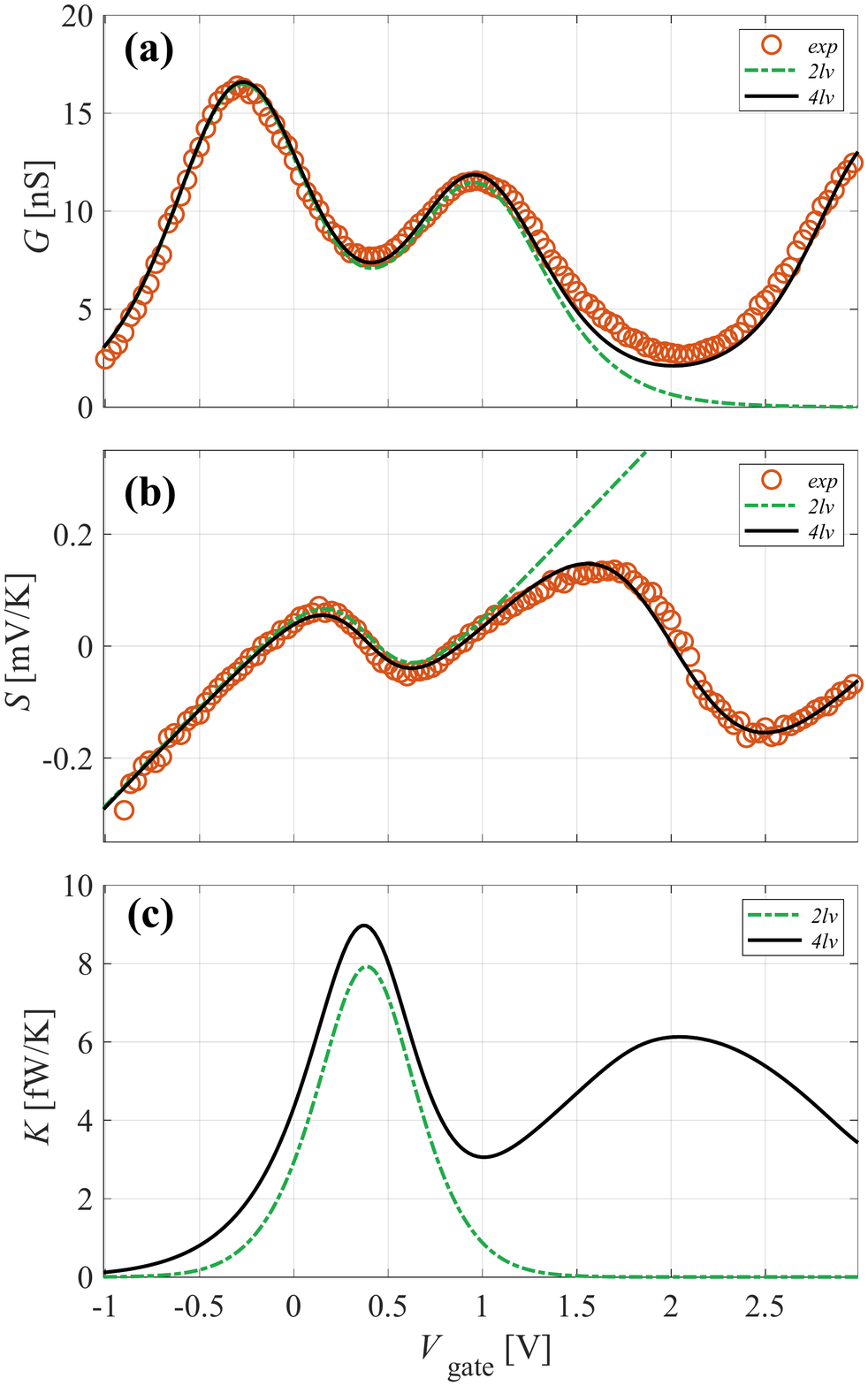}
	\caption{Theoretical fit of the experimental data and modeling of electronic thermal conductivity. Conductance (a) and thermopower (b) data shown in Fig.\ref{fig2}(e)-(f) are compared with the result of the two-level and four-level models. Little discrepancy between the two emerges in the area of interest involving the first spin-degenerate state of the QD. Differently, sizable deviations start to appear in configurations corresponding to a QD filling $N\ge 2$. The absence of deviations below the first spin-degenerate level is consistent with the $N=0$ filling. In the model, the experimental gate voltage values are converted into energy using a gate lever arm $\alpha_{\rm gate}=0.011 \pm 0.001\,{\rm meV/V}$ (see Supplementary Information for further details). (c) Electronic thermal conductance $K$ obtained using the two-level (green dashed line) and four-level (black line) model. The large discrepancy between the two models on the high end of the plotted energy range indicates only a four-level model can yield a reliable estimate of $K$ in the studied experimental configuration.}
	\label{fig3}
\end{figure}
Fig.~\ref{fig3}(a) and (b) show the experimental data (red dots) for the case $T_{\rm avg}=31.2\,{\rm K}$, together with the fitted curves (solid lines) obtained using the model ($\Delta T_{\rm fit}=2.05$ K). Different fitting models are compared: the solid black curve accounts for two spin-degenerate energy levels in the QD (here referred to as the {\em ``four-level''} model); the green dashed curve accounts for a single (spin-degenerate) energy level (referred as the {\em ``two-level''} model), as often done in the literature to model single \cite{Murphy2008,Esposito2009,Entin2010,Nakpathomkun2010,Mazza2014}  and double quantum dot system \cite{Sanchez2011,Troca2012,Jordan2013,Bhandari2018,Erdman2018}. We notice that the four-level model nicely fits the experimental data for $G$ and $S$ in the whole range of values of $V_{\rm gate}$ under consideration, while the two-level model is accurate only for values $V_{\rm gate}< 1.6\,{\rm V}$  in the conductance plot (the first two peaks in Fig. \ref{fig3}(a)) and up to $V_{\rm gate}\simeq 1\,{\rm V}$ in the thermopower plot (Fig. \ref{fig3}(b)). The theory can now be employed to calculate the electronic thermal conductance $K$ by using the fitting parameters obtained from the experimental data $G(V_{\rm gate})$ and $S(V_{\rm gate})$ (see Supplementary Information for details about values obtained for the parameters). The resulting $K(V_{\rm gate})$ curves for the two models are plotted in Fig.~\ref{fig3}(c).
Remarkably, the curve relative to the two-level model departs even in the first peak from the curve of the four-level model. This is due to the fact that K, by definition, is the ratio between the heat current and the temperature difference in open circuit conditions, i.e. when $I_{\rm DS}$ is zero. In a sequential single-level model, all electrons tunnel through the QD at the same energy, thus the same amount of heat is transferred to/from a given reservoirs in each tunneling event. If $I_{\rm DS}=0$, there is not net transfer of charge, thus there is no net transfer of heat. This implies K=0, regardless of the gate voltage. Conversely, if there are two or more energy levels, electrons can tunnel at different energies, and this possibility allows for a net heat transfer at zero charge current~\cite{Zianni2008,Erdman17}. Therefore, as opposed to G and S, the value of K in the sequential regime is fully determined by the multilevel structure of the QD. It is thus crucial to employ the four-level model to estimate it properly.

\begin{figure}[tb!]
\centering
\includegraphics[width=\linewidth]{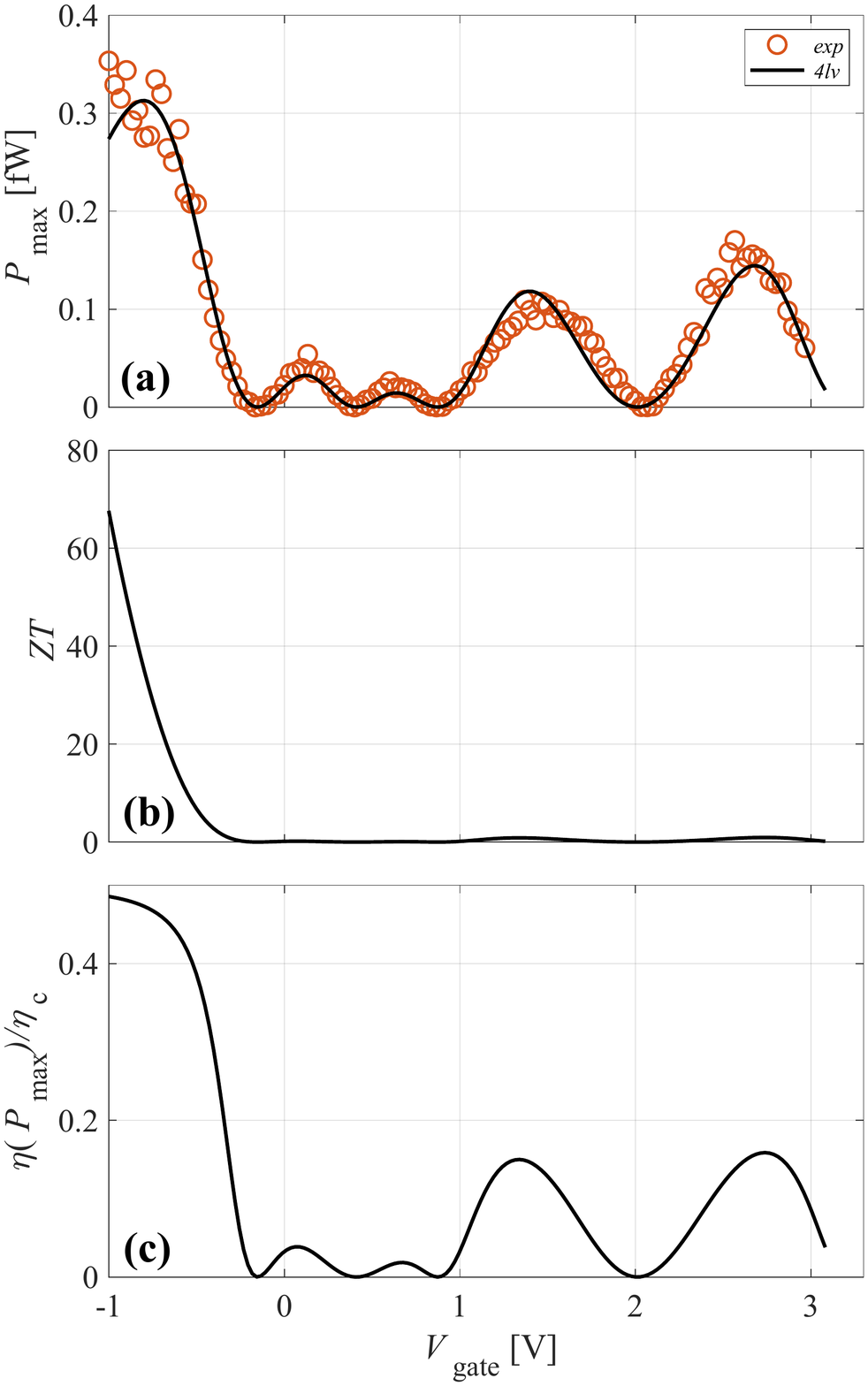}
\caption{(a) Maximum Power extracted from the system as a function of the gate voltage. The experimental data (red dots) and the theoretical prediction (black curve) are computed as $GS^2\Delta T^2/4$ using respectively the experimental and theoretical values of $G$ and $S$. (b) Electronic thermoelectric figure of merit dependence on the applied gate voltage, computed using the extrapolated curves for $G$, $S$ and $K$ (Fig. \ref{fig3}). (c) Electronic efficiency at maximum power in units of Carnot efficiency $\eta_\text{C}$. }
\label{fig4}
\end{figure}

Using the three transport coefficients $G$, $S$ and $K$ one can evaluate the performance of our QD system when operated as a heat engine. Indeed, the electric response to a temperature difference can be used to drive a current against a voltage bias, effectively extracting work from the system. The efficiency of a heat engine is then defined as the ratio between the extracted work and the heat provided by the hot reservoir. At fixed $V_{\rm gate}$ and temperature difference, we define $P_{\rm max}$ as the maximum power that can be extracted by optimizing over the applied $V_{\rm bias}$. The corresponding efficiency will then be the efficiency at maximum power $\eta(P_{\rm max}$). It can be shown\cite{Benenti2017}) that $P_{\rm max}$ also corresponds to the maximum power that can be performed on a variable load resistance in series with the QD. However, as shown in Ref.\cite{Benenti2017}, within the linear response regime both $P_{\rm max}$ and $\eta(P_{\rm max})$ can be computed simply from the knowledge of the transport coefficients G, S and K previously shown, i.e. $P_{\rm max} = GS^2 \Delta T^2/4$ and $\eta(P_{\rm max}) = \eta_\text{C}/2\times ZT/(ZT+2)$, where $ZT=GS^2T/K$ is the electronic figure of merit. The maximum power is plotted in  Fig.~\ref{fig4}(a) as a function of the gate voltage $V_{\rm gate}$ using the theoretical (solid curve) and experimental (red circles) values of $G$ and $S$ together with the fitted value of $\Delta T_{\rm fit}$. The experimental data (available down to $V_{\rm gate} \simeq -1\,{\rm V}$) is well reproduced by the four-level model which presents several peaks, the largest one corresponding to $P_{\rm max}=0.37\,{\rm fW}$ occurring at $V_{\rm gate}\simeq -0.78\,{\rm V}$. Remarkably, the corresponding value of the electronic figure of merit [plotted in Fig.~\ref{fig4}(b)] is $ZT\approx 35$. $ZT$ displays a fast increase as $V_{\rm gate}$ decreases below the first conductance peak, as expected for a single-level QD. This is due to the fact that there are no levels below the first conductance peak, so that our QD system approximately satisfies the requirement for achieving Carnot efficiency~\cite{Mahan96}. Indeed, the electronic efficiency at maximum power, plotted in Fig.~\ref{fig4}(c), takes the value $\eta(P_{\rm max}) = 0.47 \eta_{\rm C}$ (where $\eta_\text{C}$ is Carnot's efficency) for $V_{\rm gate}\simeq -0.78\,{\rm V}$, implying that our QD system can be operated at an electronic efficiency at maximum power very close to the Curzon-Ahlborn's linear response upper bound $\eta_{\rm CA} = \eta_{\rm C}/2$. It is worth observing that the electronic efficiency at maximum power is computed from the knowledge of ZT~\cite{Benenti2017}.
Interestingly, Fig.~\ref{fig4}(c) shows that the electronic efficiency at maximum power roughly behaves as the maximum power [Fig.~\ref{fig4}(a)], implying that both the maximum power and corresponding efficiency can be simultaneously maximized.

In conclusion, we have explored TE phenomena in InAs/InP NW QD-based devices at high temperature in the linear regime. Experimental data were analyzed using a multi-level model based on the resolution of a master equation which allowed us to compute the electronic thermal conductance of the system. This, in combination with the experimental data of conductance and thermopower, allowed us to estimate the electronic thermoelectric figure of merit $ZT$ and the electronic efficiency at maximum power of our thermoelectric engine. We find that the ideal Curzon-Ahlborn's upper bound is nearly attained, and that a figure of merit $ZT\approx35$ is reached while extracting the maximum power from the system. This study demonstrates the full electrostatic control of the heat engine features of a thermally biased NW QD operating in high temperature regimes. Our results shed light on the operation of few level thermoelectric engines, a key issue  for the physics and technology addressing heat and charge transport mediated by single carrier. The electronic ZT, which neglects the contribution of the phonons to the thermal conductance, is a sound characterization of the electronic properties of the device which allows to compare the electronic performance of different thermoelectric materials. Furthermore, it is directly related to the efficiency of non-equilibrum devices, such as solars cells which aim at recovering the energy of out-of-equilibrium "hot carriers" excited by light~\cite{Limpert2015,Limpert2017}. Possible applications include on-chip cooling, energy harvesting on cryogenic platforms and nanoscale thermometry.

We thank R. Fazio for helpul discussions and H. Courtois for carefully reading the manuscript. This work has also been supported by the SNS-WIS joint lab QUANTRA, by the SUPERTOP project, QUANTERA ERA-NET Cofound in Quantum Technologies and by the CNR-CONICET cooperation programme Energy conversion in quantum, nanoscale, hybrid devices. 

{\bf Materials and Methods.} InAs/InP heterostructured NWs were fabricated using a chemical beam epitaxy process seeded by metallic nanoparticles obtained from thermal dewetting of a Au thin film~\cite{Gomes2015}. Growth was performed at 420 $\pm$ 10 $\degree$C using trimethylindium (TMIn), tert-butylarsine (TBA), and tributylphosphine (TBP). The TBAs and TBP are thermally cracked at around 1000 $\degree$C upon entering the growth chamber, while the TMIn decomposes on the substrate surface. The metallorganic pressures were 0.3, 1.0, and 4.0 Torr for TMIn, TBAs, and TBP, respectively. These growth conditions ensure to achieve InAs/InP NW heterostructures with straight morphology, constant diameter, wurtzite crystal structure and atomically sharp interfaces~\cite{Zannier2017,Zannier2018}. InAs/InP and InP/InAs interfaces where realized without any interruption by switching the group V precursors. The position of the dot inside the NW is determined based on its average distance from the Au nanoparticle, as measured from an ensemble of wires with STEM. This leads to a typical alignment error of $\pm$50 nm, based on NW imaging statistics and on alignment errors during the lithographic process. Ohmic contacts are obtained by thermal evaporation of a Ti/Au (10/100 nm) bilayer, after a chemical passivation step based on a standard $\rm (NH_4)S_x$ solution. The NW voltage bias was obtained using the auxiliary output of a Stanford Research lock-in amplifier SR830 and a resistor divider. Charge current was measured using a DL1211 current preamplifier and an Agilent 34401A multimeter. Each of the thermometers was excited by an AC current of $1\,{\rm \mu A}$, obtained using the sinusoidal output of a SR830 amplifier connected to a $1\,{\rm M\Omega}$ bias resistor. The local heater was supplied using a Keithley 2400 SourceMeter. The gate voltage was controlled using a Yokogawa 7651 DC source.

\section{Supplementary Information}
\begin{itemize}
	\item Experimental methods: measurement setup, thermometers and heaters calibration, device characterization;
	\item Details about the theoretical models involved in this study;
\end{itemize}

\bibliography{refs}
\end{document}